\documentclass[aps,pra,twocolumn,amsmath,amssymb,groupedaddress,longbibliography]{revtex4-1}
\usepackage{graphicx}
\usepackage{dcolumn}
\usepackage{CJK}
\usepackage{bm}
\usepackage[dvipdfm, pdfstartview=FitH, CJKbookmarks=true, bookmarksnumbered=true, bookmarksopen=true, colorlinks=true, pdfborder=001, citecolor=blue, linkcolor=blue, linktocpage=true] {hyperref}

\begin{document}

\title{Dispersion Suppressed Topological Thouless Pumping}

\author{Shi Hu$^{1,2}$}

\author{Yongguan Ke$^{1}$}

\author{Yuangang Deng$^{1}$}

\author{Chaohong Lee$^{1,2,3}$}
\altaffiliation{Email: lichaoh2@mail.sysu.edu.cn}

\affiliation{$^{1}$Laboratory of Quantum Engineering and Quantum Metrology, School of Physics and Astronomy, Sun Yat-Sen University (Zhuhai Campus), Zhuhai 519082, China}

\affiliation{$^{2}$State Key Laboratory of Optoelectronic Materials and Technologies, Sun Yat-Sen University (Guangzhou Campus), Guangzhou 510275, China}

\affiliation{$^{3}$Synergetic Innovation Center for Quantum Effects and Applications, Hunan Normal University, Changsha 410081, China}

\begin{abstract}
  In Thouless pumping, although non-flat band has no effects on the quantization of particle transport, it induces wave-packet dispersion which hinders the practical applications of Thouless pumping.
  Indeed, we find that the dispersion mainly arises from the dynamical phase difference between individual Bloch states.
  Here we propose two efficient schemes to suppress the dispersion in Thouless pumping: a re-localization echo protocol and a high-order tunneling suppression protocol.
  In the re-localization echo protocol, we reverse the Hamiltonian in the second pumping cycle to cancel the dynamical phase difference arising from non-flat band, so that the dispersed wave-packet becomes re-localized.
  In the high-order tunneling suppression protocol, we modulate the nearest-neighbor tunneling to make the Bloch band more flat and suppress the high-order tunneling which causes wave-packet dispersion.
  Our study paves a way toward the dispersionless Thouless pumping for practical applications in matter transport, state transfer and quantum communication.
\end{abstract}

\date{\today}
\maketitle

\section{Introduction\label{Sec1}}
Topological Thouless pumping~\cite{DJThoulessPRB1983}, a quantized transport in a one-dimensional cyclically modulated periodic potential, has attracted a great attention.
In addition to the electronic system~\cite{DJThoulessPRB1983}, ultracold atomic system~\cite{LWangPRL2013} and photonic waveguide array~\cite{YKeLPR2016} have been proposed to implement Thouless pumping.
Recently, Thouless pumping has been experimentally demonstrated via cold atoms in modulated optical lattices~\cite{MLohseNP2016,SNakajimaNP2016}.
Moreover, topological Thouless pumping has been widely studied in interacting systems~\cite{JTangpanitanonPRL2016,YKePRA2017,AHaywardPRB2018}, high-dimensional system~\cite{MLohseNature2018}, and Floquet systems~\cite{IMartinPRX2017,PWeinbergPR2017,MHKolodrubetzPRL2018}.

The realization of Thouless pumping has to satisfy two key conditions: (i) the system undergoes adiabatic evolution, and (ii) the input state uniformly fills the evolved Bloch band.
The non-adiabatic effects in Thouless pumping have been well understood~\cite{WMaPRL2018,LPriviteraPRL2018,YKunoarXiv2018}.
Generally, the input state can be chosen as a Wannier state, which is strictly localized for an ideal flat band. 
During the pumping process, the input Wannier state acquires both Berry phase and dynamical phase, which shift its center position and change its spatial distribution~\cite{YKeLPR2016,YKePRA2017}.
Although a non-flat band does not affect the position shift, it induces wave-packet dispersion even under adiabatic evolution.
Due to the dispersion, large-size systems are needed to avoid boundary effects and it becomes more difficult to determine the wave-packet center.
Moreover, the dispersion will hinder the practical applications of Thouless pumping in matter transport, state transfer and quantum communication~\cite{NLang2017,CDlaskaQSC2017,ABRedondoScience2018}.
Although a flat band may suppress the dispersion, it is still a challenge to keep the band flat during the whole process.
Thus it is important to find efficient approaches to suppress the dispersion.

In this article, we present two protocols to suppress the dispersion during Thouless pumping: (i) the re-localization echo protocol and (ii) the high-order tunneling suppression protocol.
In the first protocol, the Hamiltonian is changed from $\hat H(t)$ in the first pumping cycle to $-\hat H(t)$ in the second pumping cycle.
Since the dynamical phases of individual Bloch states are opposite in the two cycles, the final dynamical phases vanish and the dispersed wave-packet becomes re-localized in the second cycle.
This re-localization echo protocol is similar to the well-developed spin echo technique~\cite{ELHahnPR1950,LMKVandersypenRMP2004,MAtalaNP2013,
DSuterRMP2016,MGarttnerNP2017}.
In the second protocol, we modulate the nearest-neighbor tunneling to switch off the high-order resonant tunneling, which is the main source of dispersion.
Actually, such a dispersion suppression attributes to the tunneling modulation makes the band more flat.

\section{Quantized mean position shift and variation of dispersion width\label{Sec2}}
\subsection{Model\label{Sec21}}
We consider the generalized commensurate Aubry-Andr\'{e}-Harper (AAH) model~\cite{PGHarperPPSA1955,SAubryAIPS1980,LLangPRL2012,YKeLPR2016},
\begin{eqnarray}\label{Eq.Ham}
 \hat{H}(t)=\sum_{j}\Big(J_{j}(t)c_{j}^{\dag}c_{j+1}
 +{\rm H.c.}\Big)
 +\sum_{j}V_{j}(t)c_{j}^{\dag}c_{j},
\end{eqnarray}
with \emph{L} \emph{q}-site cells (the total sites is $N=qL$).
Here $c_{j}^{\dag}$ ($c_{j}$) are creation (annihilation) operators for the \emph{j}-th site.
We alternately denote the $(3l-2)$-th, $(3l-1)$-th and $(3l)$-th sites as \emph{A}, \emph{B} and \emph{C}.
Their corresponding on-site energies are respectively $V_{A}$, $V_{B}$ and $V_{C}$, see Fig.~\ref{Suppressed}.
The on-site energies are modulated according to $V_{j}(t)=V_{0}\cos(2\pi\beta j+\phi(t))$ with the modulation amplitude $V_{0}$ and the rational parameter $\beta=p/q$ (where \emph{p} and \emph{q} are coprime numbers).
For the nearest-neighbor tunneling strength, we set $J_{j}(t)=-J$ for the first protocol and $J_{j}(t)=-J\sin(2\pi\beta j+\phi(t))$ for the second protocol.
In our calculations, $\beta=p/q=1/3$, and the phase $\phi(t)$ is adiabatically swept according to $\phi(t)=\omega t+\phi_{0}$. Here, $\omega$ is the ramping speed, $\phi_{0}$  is the initial modulation phase and $T=2\pi/\omega$ is the pumping period.

The topological features of a Thouless pumping can be characterized by the Chern number,
\begin{eqnarray}\label{Eq.Chern number}
 C_{m}=\frac{1}{2\pi}\int^{\pi/q}_{-\pi/q}dk\int^{T}_{0}dt F_{m}(k,t),
\end{eqnarray}
which is defined within the Brillouin-like zone $(-\pi/q<k\leq\pi/q, 0<t\leq T)$~\cite{DJThoulessPRB1983}.
Here,
$F_{m}(k,t)
=i(\langle\partial_{t}u_{m}|\partial_{k}u_{m}\rangle
-\langle\partial_{k}u_{m}|\partial_{t}u_{m}\rangle)$ is the Berry curvature and $|u_{m}(k,t)\rangle=\frac{1}{\sqrt{L}}
\sum_{j}u_{m,j}(k,t)c_{j}^{\dagger}|0\rangle$,$u_{m,j}=u_{m,j+q}$, is the periodic part of the Bloch state $|\psi_{m}(k,t)\rangle$.
%
\subsection{Relation between mean position shift and Chern number\label{Sec22}}
At the initial time $t=0$, the input state for performing Thouless pumping is chosen as the Wannier state~\cite{GHWannierPR1937}
\begin{eqnarray}\label{Eq.Wannier state}
|W_{m}(R,0)\rangle
&=&\frac{1}{\sqrt{L}}\sum_{k}e^{-ikqR}|\psi_{m}(k,0)\rangle\cr\cr
&=&\frac{1}{L}\sum_{k,j}e^{-ikqR}e^{ikj}u_{m,j}(k,0)
c_{j}^{\dagger}|0\rangle,
\end{eqnarray}
where $R$ is the cell index and \emph{m} denotes the band index.
Due to the freedom in choosing the phase of Bloch states, $e^{i\theta(k)}|\psi_{m}(k)\rangle$, the Wannier state is arbitrary.
Fortunately, by minimizing the spread function, $\Omega=\langle\hat{X}^{2}\rangle-\langle\hat{X}\rangle^{2}$, one can get the unique maximally localized Wannier state (MLWS)~\cite{NMarzariPRB1997,NMarzariRMP2012,YKePRA2017}.
Here, the position operator is defined as $\hat{X}=\sum_{j=1}^{qL}j\hat{n}_{j}$ with $\hat{n}_{j}=c_{j}^{\dag}c_{j}$.

Below we calculate the mean position shift in a pumping cycle.
The mean position at $t=0$ is given as (see Appendix A)
\begin{eqnarray}\label{Eq.Xmean0}
\langle\hat{X}_{m}(0)\rangle
=qR+\frac{1}{L}\sum_{k}
\langle u_{m}(k,0)|
i\frac{\partial}{\partial k}|u_{m}(k,0)\rangle.
\end{eqnarray}
The time-evolution is governed by the time-dependent Schrodinger equation,$i\hbar\frac{\partial}{\partial t}|\varphi(t)\rangle
=\hat{H}(t)|\varphi(t)\rangle$.
During the pumping process, as the band gap is never closed, the particle will stay in the initial band under adiabatic evolution.
The wave function $|\varphi(t)\rangle$ at time \emph{t} can be expanded in the basis of instantaneous Bloch states of the \emph{m}-th band,
\begin{eqnarray}\label{Eq.varphit}
|\varphi(t)\rangle=
\sum_{k}\exp\Big(-\frac{i}{\hbar}\int_{0}^{t}dt'E_{m}(k,t')\Big)
g_{k}(t)|\psi_{m}(k,t)\rangle,\cr\cr
\end{eqnarray}
where the coefficients $g_{k}(t)$ satisfy
\begin{eqnarray}\label{Eq.partial t gkt}
\frac{\partial}{\partial t}g_{k}(t)=
-\sum_{k'}g_{k'}(t)\langle\psi_{m}(k,t)|
\frac{\partial}{\partial t}|\psi_{m}(k',t)\rangle\cr\cr
\times{\rm exp}\Big(-\frac{i}{\hbar}\int_{0}^{t}dt'
[E_{m}(k',t')-E_{m}(k,t')]\Big),
\end{eqnarray}
with $g_{k}(0)=\frac{1}{\sqrt{L}}e^{-ikqR}$.
The instantaneous Bloch states are also eigenstates of the translation operator $\hat{T}$
\begin{eqnarray}\label{Eq.Tpsi}
\hat{T}|\psi_{m}(k',t)\rangle=e^{-ik'q}|\psi_{m}(k',t)\rangle.
\end{eqnarray}
Here $\hat{T}$ also satisfies $\hat{T}c_{j}^{\dagger}|0\rangle=c_{j+q}^{\dagger}|0\rangle$.
By differentiating Eq.~\eqref{Eq.Tpsi} and taking scalar product with $\langle\psi_{m}(k,t)|$, we get
\begin{eqnarray}
(e^{-ik'q}-e^{-ikq})
\langle\psi_{m}(k,t)|\frac{\partial}{\partial t}|\psi_{m}(k',t)\rangle=0.
\end{eqnarray}
It means that
$\langle\psi_{m}(k,t)|\frac{\partial}{\partial t}|\psi_{m}(k',t)\rangle
=\delta_{k,k'}\langle\psi_{m}(k,t)|\frac{\partial}{\partial t}|\psi_{m}(k,t)\rangle$.
So the coefficients $g_{k}(t)$ are given as
\begin{eqnarray}\label{Eq.gkt}
&&g_{k}(t)
=g_{k}(0){\rm exp}\Big(-\int_{0}^{t}dt'
\langle\psi_{m}(k,t')|\frac{\partial}{\partial t'}
|\psi_{m}(k,t')\Big)\cr\cr
&&~~~~=g_{k}(0){\rm exp}\Big(-\int_{0}^{t}dt'
\langle u_{m}(k,t')|\frac{\partial}{\partial t'}|u_{m}(k,t')\Big).
\end{eqnarray}
After a pumping cycle ($t=T$), the wave function reads
\begin{eqnarray}\label{Eq.WmRT}
|W_{m}(R,T)\rangle=
\frac{1}{\sqrt{L}}\sum_{k}e^{-ikqR}e^{i\gamma(k)}|\psi_{m}(k,0)\rangle,
\end{eqnarray}
with
\begin{eqnarray}\label{Eq.gammak}
\gamma(k)&=&\int_{0}^{T}
\big(\langle u_{m}(k,t)|i\frac{\partial}{\partial t}|u_{m}(k,t)\rangle
-E_{m}(k,t)\big)dt \nonumber \\
&=&\gamma_b(k)+\gamma_d(k).
\end{eqnarray}
Here, we respectively denote $\gamma_b$ and $\gamma_d$ as the Berry phase and the dynamical phase (here and hereafter we set $\hbar=1$).
The mean position at $t=T$ is given as
\begin{eqnarray}\label{Eq.XmeanT}
\langle\hat{X}_{m}(T)\rangle
=\langle\hat{X}_{m}(0)\rangle
-\frac{1}{L}\sum_{k}\frac{\partial}{\partial k}\gamma(k).
\end{eqnarray}
\begin{figure}[!htp]
\includegraphics[width=\columnwidth]{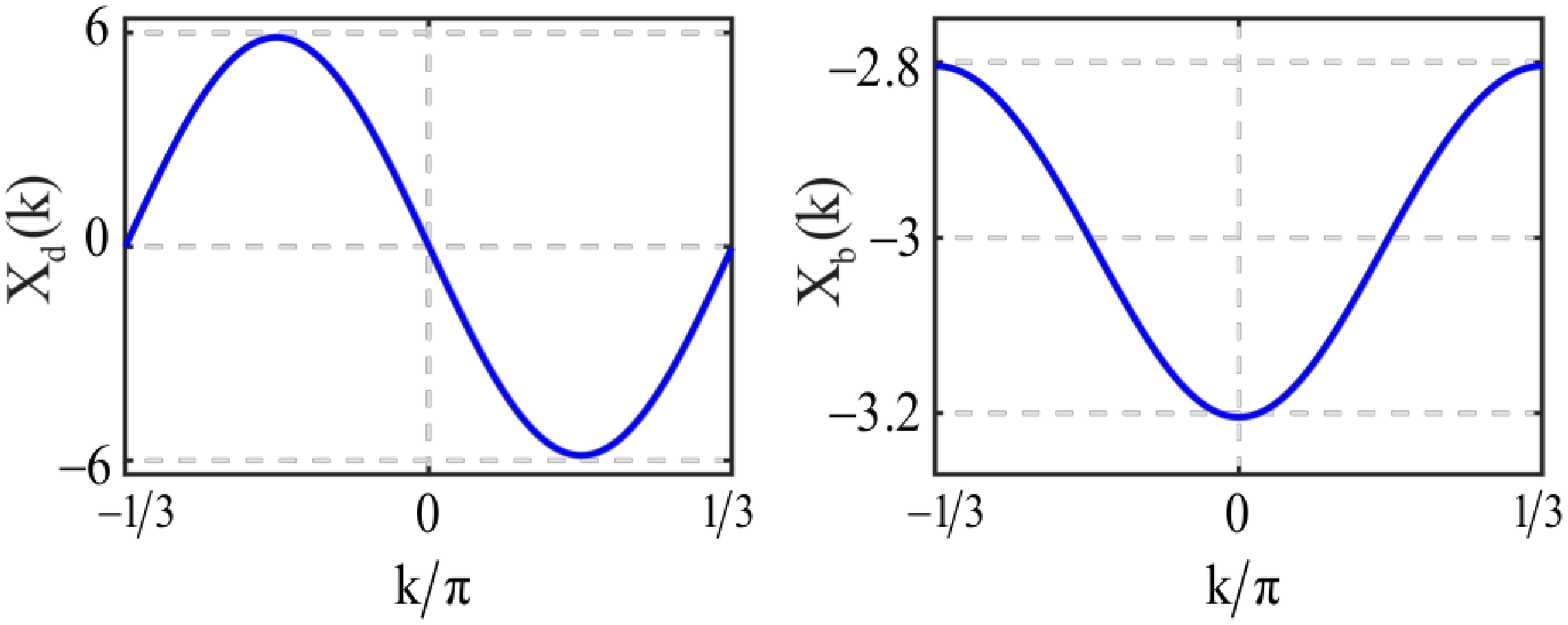}%
\caption{\label{MeanShift}(color online).
$X_{d}(k)$ (left) and $X_{b}(k)$ (right) for the highest band.
Here we chose $J_{j}(t)=-J$ and the parameters are set as $N=45$, $J=1$, $V_{0}=30$, $\phi_{0}=0$ and $\omega=0.01$.
  }
\end{figure}
In the limit of large \emph{L}, one can use the form of continuous integral to replace the summation over the quasi-momentum \emph{k}, and the mean position shift in one pumping cycle is given as
\begin{eqnarray}\label{Eq.DeltaP}
\Delta P
&=&\langle\hat{X}_{m}(T)\rangle-\langle\hat{X}_{m}(0)\rangle\cr\cr
&=&\frac{q}{2\pi}\int_{-\pi/q}^{\pi/q}X_{b}(k)dk
=qC_{m}.
\end{eqnarray}
This relationship depends only on the band topology~\cite{RDKingPRB1993,DXiaoRMP2010} and it is independent of the evolution details.
Since $E_{m}(k,t)$ is periodic with the period $2\pi/q$, the term $\frac{q}{2\pi}\int_{-\pi/q}^{\pi/q}\frac{\partial}{\partial k}
\int_{0}^{T}E_{m}(k,t)dtdk$ vanishes.
In Fig.~\ref{MeanShift}, we plot $X_{d}(k)=-\frac{\partial}{\partial k}\gamma_{d}(k)$ and $X_{b}(k)=-\frac{\partial}{\partial k}\gamma_{b}(k)$ versus the quasi-momentum \emph{k}.
It is shown that the average value of $X_{d}(k)$ vanishes and the one of $X_{b}(k)$ is $qC_{m}$.
This well agrees with the analytical value given by Eq.~\eqref{Eq.DeltaP}.
%
\subsection{Variation of dispersion width during Thouless pumping \label{Sec23}}
\begin{figure}[!htp]
\includegraphics[width=0.5\columnwidth]{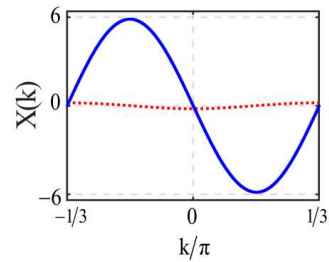}%
\caption{\label{DispersionWide}(color online).
$X_{d}(k)$ ( blue solid line) and $\xi(k)$ (red dotted line) for the highest band.
The parameters are set as the ones for Fig.~\ref{MeanShift}.
These parameters correspond to strong diagonal modulation with $|J/V_{0}|=1/30$
}
\end{figure}

However, the non-flat band makes the
dynamical phases of individual Bloch states different and this difference makes the input MLWS dispersed.
This dispersion can be characterized by the dispersion width $D_{W}=\sqrt{\Omega}$, which is the square root of the spread function $\Omega$.
To get the variation of $D_{W}$, we analyze the spread functional first.
For the isolated band we consider, the spread functional of the Wannier states in the \emph{m}-th band can be written as
\begin{eqnarray}\label{Eq.Omega}
\Omega(t)=\langle\hat{X}^{2}\rangle-
\langle\hat{X}\rangle^{2},
\end{eqnarray}
with $\langle\hat{X}^{2}\rangle=\langle W_{m}(0,t)|\hat{X}^{2}|W_{m}(0,t)\rangle$ and
$\langle\hat{X}\rangle=\langle W_{m}(0,t)|\hat{X}|W_{m}(0,t)\rangle$.
It can be decomposed as two terms, $\Omega=\Omega_{I}+\Omega_{D}$, with
\begin{eqnarray}\label{Eq.OmegaI}
&&\Omega_{I}=\sum_{m'\neq m, R}
|\langle W_{m'}(R,t)|\hat{X}|W_{m}(0,t)\rangle|^{2},\cr\cr
&&\Omega_{D}=\sum_{R\neq 0}
|\langle W_{m}(R,t)|\hat{X}|W_{m}(0,t)\rangle|^{2}.
\end{eqnarray}
Here, $\Omega_{I}$ is gauge invariant and therefore doesn't change during the whole pumping procedure, and $\Omega_{D}$ is zero for the MLWSs~\cite{NMarzariPRB1997,NMarzariRMP2012,YKePRA2017}.
Therefore the dispersion after one pumping cycle is determined by (see
Appendix B)
\begin{eqnarray}\label{Eq.OmegaDT}
\Omega_{D}(T)
=\frac{1}{L}\sum_{k}\Big(-\frac{\partial}{\partial k}\gamma(k)
+\frac{1}{L}\sum_{k}\frac{\partial}{\partial k}\gamma(k)\Big)^{2}.
\end{eqnarray}
In the limit of large \emph{L}, we can use the form of continuous integral to replace the summation over quasi-momentum \emph{k} and obtain
\begin{eqnarray}\label{Eq.OmegaDT1}
\Omega_{D}(T)
=\frac{q}{2\pi}\int_{-\pi/q}^{\pi/q}
\Big(X_{d}(k)+X_{b}(k)-qC_{m}\Big)^{2}dk.
\end{eqnarray}
Actually, the quantized transport $qC_{m}$ is just the average value of $X_{b}(k)$ over the Brillouin-like zone.
The term $X_{d}(k)$ arises from the non-uniformly dynamical phases accumulated in the pumping process, which gradually induce dispersion, and so that more significant dispersion appears for slower modulation.
In contrast, the term $\xi(k)=X_{b}(k)-qC_{m}$ does not depend on the evolution time.
In Fig.~\ref{DispersionWide}, we plot $X_{d}(k)$ and $\xi(k)=X_{b}(k)-qC_{m}$ versus the quasi-momentum \emph{k}.
It clearly shows that $X_{d}(k)$ is the main source of dispersion under strong diagonal modulation (i.e. $|J/V_{0}|\ll1$).

%
\section{Dispersion suppressed topological Thouless pumping\label{Sec3}}
\subsection{Re-localization echo protocol\label{Sec31}}
\begin{figure}[!htp]
\includegraphics[width=\columnwidth]{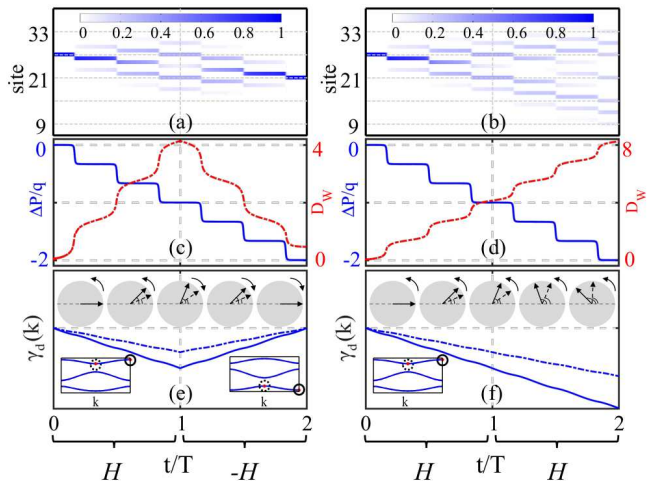}
\caption{\label{Echo}(color online).
Thouless pumping during two pumping cycles: (i) the re-localization echo protocol (left column), and (ii) the traditional protocol (right column).
(a, b): the density distribution $\langle\hat{n}_{j}\rangle$ versus $t/T$. (c, d): the mean position shift $\Delta P$ (blue solid line) and the dispersion width $D_{W}$ (red dashed line) versus $t/T$.
(e, f): the dynamical phase $\gamma_d(k)$ versus $t/T$.
The insets in (e, f) illustrate the energy bands.
The solid (dashed) arrows and blue solid (dashed) lines correspond to the points in solid (dashed) circles.
The parameters are set as the ones for Fig.~\ref{MeanShift}.
}
\end{figure}

In this protocol, we discuss how to suppress the dispersion via cancelling the dynamical phases.
The system evolves under $\hat{H}(t)$ in the first cycle and then $-\hat{H}(t)$ in the second cycle.
At the beginning, each individual Bloch state $|\psi_m(k,0)\rangle$ has no dynamical phase, see the horizontal arrows in Fig.~\ref{Echo}~(e).
The dynamical phase difference increases with the evolution time during the first cycle, see the anticlockwise-rotating arrows with different frequencies.
In the second cycle, since the Hamiltonian changes from $\hat{H}(t)$ to $-\hat{H}(t)$, the eigenstate index changes from $m$ in the first cycle to $m'=q+1-m$ in the second cycle, and the energy bands become reversed, see the insets in Fig.~\ref{Echo}~(e).
This means that the corresponding Wannier state also changes from filling the $m$-th band of $H(t)$ to filling the $m'$-th band of $-H(t)$.
As $|\psi_{m}(k,t)\rangle$ and $|\psi_{m'}(k,T+t)\rangle$ represent the same quantum state, this ensures that the Berry phases accumulated in the two cycles are equal.
Thus the Chern numbers for the two bands are exactly the same.
Since the mean position shift $\Delta P$ only depends on the Chern number, it takes the same value in each cycle.

Due to the band inversion, the Bloch states $|\psi_{m}(k,t)\rangle$ and $|\psi_{m'}(k,T+t)\rangle$ evolve with opposite energy.
As a result, the dynamical phase for the second cycle $\gamma_{d}^{(2)}$ is just the opposite to the one for the first cycle $\gamma_{d}^{(1)}$, that is,
$\gamma_{d}^{(2)}
=-\int_{T}^{2T}\langle\psi_{m'}(k,t)|\hat{H}(t) |\psi_{m'}(k,t)\rangle dt
=\int_{0}^{T}\langle\psi_{m}(k,t)|\hat{H}(t) |\psi_{m}(k,t)\rangle dt
=-\gamma_{d}^{(1)}$.
Consequently, the dynamical phases accumulated in the second cycle cancel the ones accumulated in the first cycle, see the clockwise-rotating arrows back to the initial direction in  Fig.~\ref{Echo}~(e).
The blue solid (dashed) lines respectively denote the dynamical phases of the Bloch states corresponding to the solid (dashed) circles in the bands.
In contrast, in the traditional Thouless pumping without reversing the sign of $H(t)$, the dynamical phase difference and the dispersion width will increase with the evolution time, see Fig.~\ref{Echo}~(f).

In Fig.~\ref{Echo}, we compare our re-localization echo protocol with the traditional Thouless pumping.
At the beginning, the particle stays in the 27-th site (i.e. the $C$-sublattice at the $9$-th cell), which is labeled as $|C\rangle_{9}$.
The initial state has 99.9\% projection on the MLWS for the highest band.
In the first cycle, the density distribution gradually spreads with the evolution time.
In the second cycle, for the traditional Thouless pumping, the density distribution spreads as the one in the first cycle, see Fig.~\ref{Echo}~(b).
However, for the re-localization echo protocol, the density distribution re-localizes in the second cycle and the final distribution almost recovers its initial shape, see Fig.~\ref{Echo}~(a).
The final state in the re-localization echo protocol has $98.9\%$ projection on the MLWS $|C\rangle_{7}$, but the final state in the traditional Thouless pumping has only a very small projection on $|C\rangle_{7}$.
Although the mean position shifts $\Delta P$ are almost the same (2 unit cells) for both the re-localization echo protocol [Fig.~\ref{Echo}~(c)] and the traditional Thouless pumping [Fig.~\ref{Echo}~(d)], the corresponding dispersion widths are very different.
In the re-localization echo protocol, the dispersion width $D_{W}$ (red dashed line) gradually increases in the first cycle and then gradually decreases to $0$ in the second cycle, see Fig.~\ref{Echo}~(c).
In the traditional Thouless pumping, the dispersion width $D_{W}$ (red dashed line) keeps increase with the evolution time, see Fig.~\ref{Echo}~(d).
Thus, although the mean position shifts are both determined by Chern number, the wave-packet dispersion is strongly suppressed by the re-localization echo protocol.

\subsection{High-order tunneling suppression protocol\label{Sec32}}
\begin{figure}
\centering
\includegraphics[width=\columnwidth]{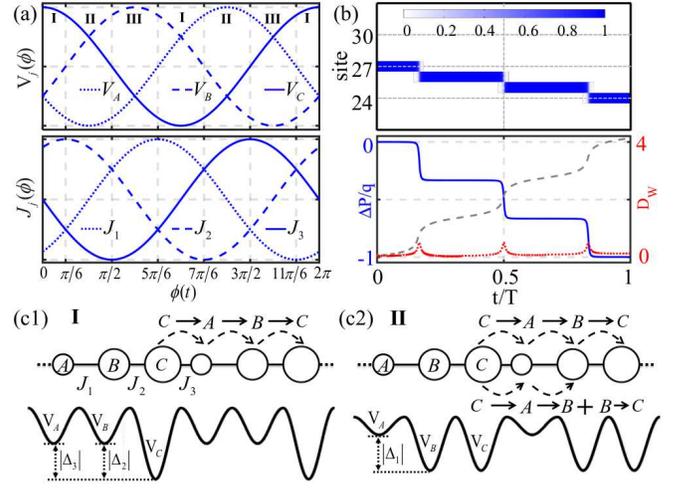}%
\caption{\label{Suppressed}(color online).
Dispersion suppressed Thouless pumping under modulation.
(a) Modulated on-site energies and nearest-neighbor tunneling terms.
(b) Top: the density distribution $\langle \hat{n}_{j}\rangle$. Bottom: the mean position shift $\Delta P$ and the dispersion wide $D_{W}$.
The blue solid and red dashed-dot lines denote $\Delta P$ and $D_{W}$, respectively. The gray dashed line corresponds to the system without tunneling modulation.
(c) High-order resonant tunneling processes in regions (I) and (II).
The other parameters are set as the ones for Fig.~\ref{MeanShift}.
}
\end{figure}

The dispersion mechanism in Thouless pumping can also attribute to the high-order quantum tunneling.
Under strong diagonal modulation ($|J/V_{0}|\ll1$), due to the first-order resonant tunneling, the mean position shift mainly occurs around the time when the on-site potentials of neighbouring sites are equal, see Fig.~\ref{Echo}.
If the particle only jumps to the nearest neighboring sites, the wave-packet will keep localized.
However, at the same time, the high-order resonant tunneling also takes place and thus the wave-packet disperses, see Fig.~\ref{Suppressed}(c).
As a result, the dispersion width $D_{W}$ grows drastically around the resonant points, see Fig.~\ref{Echo}.
Unlike the unidirectional first-order resonant tunneling, the high-order resonant tunneling always occurs between neighbouring cells and has equal probability to spread toward opposite directions.

To describe the dispersion caused by the inter-cell tunneling, we derive the effective model via applying degenerate perturbation theory.
Under strong diagonal modulation ($|J/V_{0}|\ll1$), we treat the tunneling term
\begin{eqnarray}\label{Eq.tunneling}
\hat{V}=\sum_{j}\Big(J_{j}c_{j}^{\dag}c_{j+1}+{\rm H.c.}\Big),
\end{eqnarray}
as a perturbation to the on-site energy term
\begin{eqnarray}\label{Eq.on-site potential}
\hat{H}_{0}=\sum_{j}V_{j}c_{j}^{\dag}c_{j}.
\end{eqnarray}

\begin{figure}[!htp]
\includegraphics[width=\columnwidth]{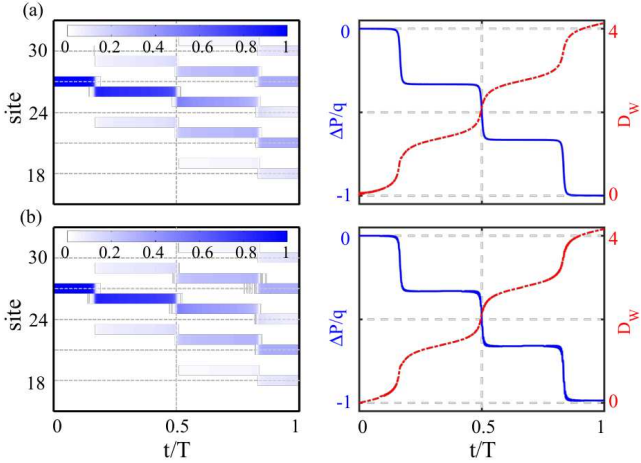}%
\caption{\label{Effective}(color online).
Thouless pumping during one pumping cycle. Left: the density distribution $<\hat{n}_{j}>$. Right: the mean position shift $\Delta P$ and the dispersion width $D_{W}$.
(a) and (b) respectively correspond to the original and effective Hamiltonians.
The parameters are set as the ones for Fig.~\ref{MeanShift}.
}
\end{figure}

Since different resonant points corresponding to different effective Hamiltonians, we decompose one pumping cycle as three regions: (I) around the first resonant point where $V_{A}=V_{B}$, (II) around the second resonant point where $V_{B}=V_{C}$, and (III) around the third resonant point where $V_{C}=V_{A}$.
The unperturbed term $\hat{H}_{0}$ has three eigenvalues, $E_{1}=V_{A}$, $E_{2}=V_{B}$, and $E_{3}=V_{C}$ each with the \emph{L}-fold degenerate eigenstates.
As an example, we calculate the effective Hamiltonian for region I, where $V_{A}$ and $V_{B}$ are far separated from $V_{C}$.
The effective Hamiltonians for regions II and III are quite similar and will be given at last.
The eigenstates of $\hat{H}_{0}$ construct two different subspaces: (1) the subspace $\mathcal{U}$ formed by $\{|A\rangle_{l},|B\rangle_{l}\}$ (with l=1,...,L), and (2) the subspace $\mathcal{V}$ formed by $\{|C\rangle_{l}\}$ (with l=1,...,L).
The projection operators on spaces $\mathcal{U}$ and $\mathcal{V}$ are respectively defined as
\begin{eqnarray}\label{Eq.projection operators}
\hat{P}_{0}&=&\sum_{l}|A\rangle_{l}\langle A|_{l}
+|B\rangle_{l}\langle B|_{l}\cr\cr
\hat{Q}_{0}&=&\sum_{l}|C\rangle_{l}\langle C|_{l}.
\end{eqnarray}
According to the Schrieffer-Wolff transformation method ~\cite{MTakahashiJPC1977,SBravyiAP2011}, the effective Hamiltonian for the subspace $\mathcal{U}$ is given as
\begin{eqnarray}\label{Eq.HeffU}
\hat{H}_{\mathcal{U}}&=&\hat{H}_{0}\hat{P}_{0}+\hat{P}_{0}\hat{V}\hat{P}_{0}
+\sum_{n=2}^{\infty}\hat{H}_{eff,n},\cr\cr
\hat{H}_{eff,n}&=&\sum_{j\geq1}b_{2j-1}
\hat{P}_{0}\tilde{S}^{2j-1}(V_{od})_{n-1}\hat{P}_{0},\cr\cr
\tilde{S}^{2j-1}(V_{od})_{n-1}
&=&\sum_{\substack{m_{1},\dots,m_{2j-1}\geq1\\m_{1}+\dots+m_{2j-1}=n-1}}
\tilde{S}_{m_{1}}\dots\tilde{S}_{m_{2j-1}}(V_{od})\cr\cr
&&
\end{eqnarray}
Here $b_{2j-1}$ are the Taylor coefficients of the function tanh$(x/2)$ and the super-operator $\tilde{S}$ describing the adjoint action of $\tilde{S}$, that is, $\tilde{S}(V_{od}) = [S, V_{od}]$.
In this article we consider the effective Hamiltonian up to third-order.
The second-order effective Hamiltonian is
\begin{eqnarray}\label{Eq.Heff2}
\hat{H}_{eff,2}=
\frac{1}{2}\hat{P}_{0}\tilde{S}_{1}(V_{od})\hat{P}_{0},
\end{eqnarray}
and the third-order effective Hamiltonian is
\begin{eqnarray}\label{Eq.Heff3}
\hat{H}_{eff,3}=
\frac{1}{2}\hat{P}_{0}\tilde{S}_{2}(V_{od})\hat{P}_{0},
\end{eqnarray}
with
\begin{eqnarray}
S_{1}&=&\mathcal{L}(V_{od}),~~
S_{2}=-\mathcal{L}([V_{d},S_{1}]),\cr\cr
V_{d}&=&\mathcal{D}(\hat{V}),~~~
V_{od}=\mathcal{O}(\hat{V}).
\end{eqnarray}
Here, the super-operators are defined as
\begin{eqnarray}\label{Eq.superoprator}
\mathcal{O}(\hat{Y})&=&P_{0}\hat{Y}Q_{0}+Q_{0}\hat{Y}P_{0},\cr\cr
\mathcal{D}(\hat{Y})&=&P_{0}\hat{Y}P_{0}+Q_{0}\hat{Y}Q_{0},\cr\cr
\mathcal{L}(\hat{Y})&=&\sum_{i,j}
\frac{\langle i|\mathcal{O}(\hat{Y})|j\rangle}{E_{i}-E_{j}}
|i\rangle\langle j|,
\end{eqnarray}
with $\{|i\rangle\}$ be an orthonormal eigenbasis of $\hat{H}_{0}$ and $\hat{H}_{0}|i\rangle=E_{i}|i\rangle$ for all \emph{i}.
We denote the eigenvalues and eigenstates belonging to the subspace $\mathcal {U}$ as $\{E_{l_0}\}$ and $|l_0\rangle$, respectively.
In this way the first-order effective Hamiltonian write as
\begin{eqnarray}\label{Eq.firstorder}
\hat{P}_{0}\hat{V}\hat{P}_{0}
=\sum_{i,j\in l_0}\langle i|\hat{V}|j\rangle
|i\rangle \langle j|,
\end{eqnarray}
the second-order effective Hamiltonian write as
\begin{eqnarray}\label{Eq.secondorder}
\hat{H}_{eff,2}
=\sum_{i,j\in l_0, m \notin l_0}
&&\Big[\frac{1}{2}\Big(\frac{1}{E_{i}-E_{m}}+\frac{1}{E_{j}-E_{m}}\Big)\times\cr\cr
&&\langle i|\hat{V}|m\rangle\langle m|\hat{V}|j\rangle |i\rangle \langle j|\Big],
\end{eqnarray}
and the third-order effective Hamiltonian write as
\begin{eqnarray}\label{Eq.thirdorder}
\hat{H}_{eff,3}
&=&\sum_{i,j\in l_{0}, m,n\notin l_{0}}
\frac{\langle j|\hat{V}|m\rangle\langle m|\hat{V}|n\rangle
\langle n|\hat{V}|i\rangle}{2(E_{i}-E_{m})(E_{i}-E_{n})}
|j\rangle\langle i|
\cr\cr
&+&\sum_{i,j\in l_{0}, m,n\notin l_{0}}
\frac{\langle i|\hat{V}|m\rangle\langle m|\hat{V}|n\rangle
\langle n|\hat{V}|j\rangle}{2(E_{i}-E_{m})(E_{i}-E_{n})}
|i\rangle\langle j|
\cr\cr
&-&\sum_{i,j,k\in l_{0}, m\notin l_{0}}
\frac{\langle k|\hat{V}|m\rangle\langle m|\hat{V}|i\rangle
\langle i|\hat{V}|j\rangle}{2(E_{i}-E_{m})(E_{j}-E_{m})}
|k\rangle\langle j|
\cr\cr
&-&\sum_{i,j,k\in l_{0}, m\notin l_{0}}
\frac{\langle j|\hat{V}|i\rangle\langle i|\hat{V}|m\rangle
\langle m|\hat{V}|k\rangle}{2(E_{i}-E_{m})(E_{j}-E_{m})}
|j\rangle\langle k|.\cr\cr
&&
\end{eqnarray}
Similarly, the effective Hamiltonian for the subspace $\mathcal{V}$ can also be obtained and the total effective Hamiltonian for region I read as
\begin{eqnarray}\label{Eq.HeffI}
\hat{H}_{\rm{I}}
&=&\sum_{l}V_{\rm{I}A}c_{l,A}^{\dagger}c_{l,A}
+V_{\rm{I}B}c_{l,B}^{\dagger}c_{l,B} +V_{\rm{I}C}c_{l,C}^{\dagger}c_{l,C}\cr\cr
&+&\Big[J_{\rm{I}}^{(1)}c_{l,A}^{\dagger}c_{l,B}
+J_{\rm{I}}^{(2)}c_{l,A}^{\dagger}c_{l-1,B}
-J_{\rm{I}}^{(3)}\Big(c_{l,A}^{\dagger}c_{l+1,A}\cr\cr
&+&c_{l,B}^{\dagger}c_{l+1,B}-2c_{l,C}^{\dagger}c_{l+1,C}\Big) +\rm{H.c.}\Big].
\end{eqnarray}
Where $V_{\rm{I}A}=V_{A}+\frac{J_{3}^{2}}{\Delta_{3}}$, $V_{\rm{I}B}=V_{B}+\frac{J_{2}^{2}}{\Delta_{2}}$, and
$V_{\rm{I}C}=V_{C}-\frac{J_{2}^{2}}{\Delta_{2}} -\frac{J_{3}^{2}}{\Delta_{3}}$ are effective on-site potentials and
$J_{\rm{I}}^{(1)}=J_{1}-\frac{J_{1}(J_{2}^{2} +J_{3}^{2})}{2\Delta_{2}\Delta_{3}}$,
$J_{\rm{I}}^{(2)}=\frac{J_{2}J_{3}}{2}(\frac{1}{\Delta_{2}} +\frac{1}{\Delta_{3}})$, and $J_{\rm{I}}^{(3)}=\frac{J_{1}J_{2}J_{3}}{2\Delta_{2}\Delta_{3}}$ are respectively the first-, second-, and third-order tunneling strengths.
The second term in $J_{\rm{I}}^{(1)}$ represents a correction from the third-order process.
Here $J_{1}$, $J_{2}$, and $J_{3}$ are respectively the tunneling strengthes between \emph{A} and \emph{B}, \emph{B} and \emph{C} in the same cell, and \emph{C} and \emph{A} in two nearest-neighbor cells.
The potential biases are denoted as
$\Delta_{1}=V_{A}-V_{B}, \Delta_{2}=V_{B}-V_{C}, \Delta_{3}=V_{A}-V_{C}$.
The effective Hamiltonian for one pumping cycle is given as
\begin{equation}\label{Eq.HT}
\hat{H}_{T}=
\begin{cases}
\hat{H}_{\rm{I}},~~\phi(t)\in[0,\pi/6)\cup[5\pi/6,7\pi/6)\cup[11\pi/6,2\pi] \\
\hat{H}_{\rm{II}},~\phi(t)\in[\pi/6,\pi/2)\cup[7\pi/6,3\pi/2) \\
\hat{H}_{\rm{III}},\phi(t)\in[\pi/2,5\pi/6)\cup[3\pi/2,11\pi/6)
\end{cases}
\end{equation}
with $\hat{H}_{\rm{I}}$ given in Eq.~\eqref{Eq.HeffI} and
\begin{eqnarray}\label{Eq.HeffII}
&&\hat{H}_{\rm{II}}
=\sum_{l}V_{\rm{II}A}c_{l,A}^{\dagger}c_{l,A}
+V_{\rm{II}B}c_{l,B}^{\dagger}c_{l,B} +V_{\rm{II}C}c_{l,C}^{\dagger}c_{l,C}\cr\cr
&&+\Big[J_{\rm{II}}^{(1)}c_{l,B}^{\dagger}c_{l,C}
+J_{\rm{II}}^{(2)}c_{l+1,B}^{\dagger}c_{l,C}
-J_{\rm{II}}^{(3)}\Big(-2c_{l,A}^{\dagger}c_{l+1,A}\cr\cr
&&+c_{l,B}^{\dagger}c_{l+1,B}+c_{l,C}^{\dagger}c_{l+1,C}\Big) +\rm{H.c.}\Big]
\cr\cr
&&\hat{H}_{\rm{III}}
=\sum_{l}V_{\rm{III}A}c_{l,A}^{\dagger}c_{l,A}
+V_{\rm{III}B}c_{l,B}^{\dagger}c_{l,B} +V_{\rm{III}C}c_{l,C}^{\dagger}c_{l,C}\cr\cr
&&+\Big[J_{\rm{III}}^{(1)}c_{l+1,A}^{\dagger}c_{l,C}
+J_{\rm{III}}^{(2)}c_{l,A}^{\dagger}c_{l,C}
-J_{\rm{III}}^{(3)}\Big(c_{l,A}^{\dagger}c_{l+1,A}\cr\cr
&&-2c_{l,B}^{\dagger}c_{l+1,B}+c_{l,C}^{\dagger}c_{l+1,C}\Big) +\rm{H.c.}\Big]
\end{eqnarray}
Where $V_{\rm{II}A}=V_{A}+\frac{J_{1}^{2}}{\Delta_{1}} +\frac{J_{3}^{2}}{\Delta_{3}}$, $V_{\rm{II}B}=V_{B}-\frac{J_{1}^{2}}{\Delta_{1}}$,
$V_{\rm{II}C}=V_{C}-\frac{J_{3}^{2}}{\Delta_{3}}$,
$J_{\rm{II}}^{(1)}=J_{2}-\frac{J_{2}(J_{1}^{2} +J_{3}^{2})}{2\Delta_{1}\Delta_{3}}$,
$J_{\rm{II}}^{(2)}=-\frac{J_{1}J_{3}}{2}(\frac{1}{\Delta_{1}} +\frac{1}{\Delta_{3}})$, $J_{\rm{II}}^{(3)}=\frac{J_{1}J_{2}J_{3}}{2\Delta_{1}\Delta_{3}}$ and
$V_{\rm{III}A}=V_{A}+\frac{J_{1}^{2}}{\Delta_{1}}$, $V_{\rm{III}B}=V_{B}-\frac{J_{1}^{2}}{\Delta_{1}} +\frac{J_{2}^{2}}{\Delta2}$,
$V_{\rm{III}C}=V_{C}-\frac{J_{2}^{2}}{\Delta_{2}}$,
$J_{\rm{III}}^{(1)}=J_{3}-\frac{J_{3}(J_{1}^{2} +J_{2}^{2})}{2\Delta1\Delta_{2}}$,
$J_{\rm{III}}^{(2)}=\frac{J_{1}J_{2}}{2}(\frac{1}{\Delta_{1}} -\frac{1}{\Delta_{2}})$, $J_{\rm{III}}^{(3)}=\frac{J_{1}J_{2}J_{3}}{2\Delta_{1}\Delta_{2}}$.
In Fig.~\ref{Effective}, we compare the Thouless pumping under the original Hamiltonian (Eq.~\eqref{Eq.Ham}) and the effective Hamiltonian $\hat{H}_{T}$ (Eq.~\eqref{Eq.HT}).
Clearly, the results obtained from the effective Hamiltonian are agree with those obtained from the original Hamiltonian.

To understand the dispersion process, we discuss how the inter-cell high-order resonant tunneling appears.
As an example, we consider a particle initially occupied the 27-th site (i.e. the sublattice \emph{C} of the 9-th cell), and the system evolves in the region (I).
Because of the large potential bias, it will stay in the sublattice \emph{C} until the tunneling strength is comparable to the potential bias.
The three-order resonant tunneling makes the particle move to the sublattice \emph{C} in adjacent cells, that is, $|\emph{C}\rangle_l \rightarrow |\emph{C}\rangle_{l+1}+|\emph{C}\rangle_{l-1}$, see Fig.~\ref{Suppressed}~(c1).
Since the leftward 
and rightward 
three-order tunneling processes have the same strength, they do not change the mean position shift and may only cause wave-packet  dispersion.
Then, according to the on-site modulation $V_{j}(t)=V_{0}\cos(2\pi\beta j+\phi(t))$, the system evolves into the region~(II).
In the vicinity of $V_B = V_C$, due to the effective first-order tunneling $J_{\rm{II}}^{(1)}\big(|B\rangle_{l}\langle C|_{l}+\rm{H.c.}\big)$, the particle jumps from the sublattice \emph{C} to its nearest-neighbor sublattice \emph{B}.
In addition to the first-order resonant tunneling $|C\rangle_{l}\rightarrow|B\rangle_{l}$, the second-order resonant tunneling $|C\rangle_{l}\rightarrow|B\rangle_{l+1}$ and $|B\rangle_{l}\rightarrow|C\rangle_{l-1}$ may occur at the same time,
which are described by $J_{\rm{II}}^{(2)}\big(|B\rangle_{l+1}\langle C|_{l}+\rm{H.c.}\big)$.
In particular, the wave-packet dispersion becomes significant because these second-order tunneling processes connect with the first-order ones, that is, $|C\rangle_{l}\rightarrow|B\rangle_{l+1}\rightarrow|C\rangle_{l+1}$ and $|C\rangle_{l}\rightarrow|B\rangle_{l}\rightarrow|C\rangle_{l-1}$, see Fig.~\ref{Suppressed}~(c2).
In the whole process, the third-order resonant tunneling always takes place, and its strength is given as $J_{\rm{II}}^{(3)}\big(-2|A\rangle_{l+1}\langle A|_{l} +|B\rangle_{l+1}\langle B|_{l} +|C\rangle_{l+1}\langle C|_{l}+\rm{H.c.}\big)$.
The time-evolutions in all regions are similar: the third-order resonant tunneling and the connected second-order-to-first-order resonant tunneling cause wave-packet dispersion, but do not affect the mean position shift.

To suppress the wave-packet dispersion, one may switch off the high-order resonant tunneling.
In general, there are several different schemes to switch off the high-order resonant tunneling.
We find that one can switch off the high-order resonant tunneling via modulating the tunneling strength as
\begin{eqnarray}\label{Eq.Jj}
J_{j}(t)=-J\sin(2\pi\beta j+\phi(t)).
\end{eqnarray}
Obviously, at the resonant point $V_{B}(t)=V_{C}(t)$, $\{J_{1}, J_{2}, J_{3}\}$ are $\{0, \sqrt{3}/2, -\sqrt{3}/2\}$, respectively.
Thus we have the high-order tunneling strengthes $J_{\rm{II}}^{(2)}=J_{\rm{II}}^{(3)}=0$.
Similarly, at other resonant points [Fig.~\ref{Suppressed}~(a)], the high-order tunneling is also switched off.

In Fig.~\ref{Suppressed}, we show the dispersion suppressed topological Thouless pumping under the tunneling modulation~\eqref{Eq.Jj}.
In the top panel of Fig.~\ref{Suppressed}~(b), we show the time-evolution of the density distribution $\langle\hat{n}_{j}\rangle$.
The initial state is $|C\rangle_{9}$ whose projection on the MLWS for the highest band is 99.9\%.
The density distribution is well localized during the whole pumping process and the final state has 99.9\% projection on the state $|C\rangle_{8}$.
In the bottom panel of Fig.~\ref{Suppressed}~(b), we show the mean position shift $\Delta P$ (blue solid line) and the dispersion width $D_{W}$ (red dashed-dot line).
The mean position shifts $-0.999$ unit cells, which is very close to the Chern number $-1$ for the highest band.
The dispersion width $D_{W}$ is very small during whole process and the sharp peaks correspond to the steps in $\Delta P$.
However, if the tunneling strength is fixed, the dispersion width $D_{W}$ will increase with time, see the gray dashed line in the bottom panel of Fig.~\ref{Suppressed}~(b).

\begin{figure}[!htp]
\includegraphics[width=\columnwidth]{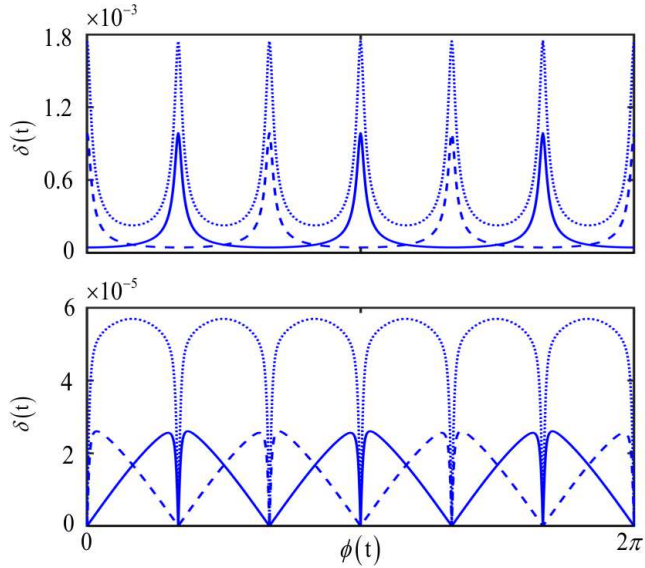}%
\caption{\label{Flatness}(color online).
Flatness ratios during one pumping cycle with fixed tunneling strength (a) and with modulated one (b).
The solid, dotted and dashed lines respectively correspond to the lowest, middle and highest bands.
The parameters are set as the ones for Fig.~\ref{MeanShift}.
}
\end{figure}

In high-order tunneling suppression protocol, we modulate the nearest-neighbor tunneling strength.
This modulation suppresses the effectively high-order tunneling processes and we find it also makes the Bloch band more flat.
The energy gap between the \emph{m}-th and (\emph{m}+1)-th bands is defined as $G_{m}=\rm{min}_{\{k\}}\left(E_{m+1,k}-E_{m,k}\right)$.
The bandwidth for \emph{m}-th band is defined as
$W_{m}=\rm{max}_{\{k\}}E_{m,k}-\rm{min}_{\{k\}}E_{m,k}$.
The flatness ratios for our three-band system are respectively given as $\delta_{1}=W_{1}/G_{1}$, $\delta_{2}=W_{2}/\rm{min}(G_{1},G_{2})$ and $\delta_{3}=W_{3}/G_{2}$.
In Fig.~\ref{Flatness}, we show the flatness ratios with respect to the time-dependent phase $\phi(t)$.

\section{Summary and Discussions\label{Sec4}}
In Thouless pumping, even under adiabatic evolution, the non-flat band will bring non-uniformly dynamical phases and then induce wave-packet dispersion.
We have put forward two protocols to achieve dispersion suppressed Thouless pumping.
In re-localization echo protocol, the initial MLWS will expand in the first cycle and it gradually localizes during the second cycle via reversing the Hamiltonian.
In high-order tunneling suppression protocol, by modulating the nearest-neighbor tunneling strength, the wave-packet is almost dispersionless during the whole pumping process.
In both protocols, the particle transports are well consistent with the quantized ones given by the Chern number and the final wave-packets almost perfectly return to their input shapes.
Our studies pave a way toward implementing long-distance dispersionless Thouless pumping for practical applications.

We also note that non-adiabatic charge pumping has been widely studied such as in discrete time crystals~\cite{KMizutaPRL2018}.
Recently, a non-adiabatic quantized charge pumping has been demonstrated in two-dimensional periodically driven quantum systems with spatial disorders~\cite{PTitumPRX2016}.
Hence, it would be interesting to achieve dispersionless charge pumping in a non-adiabatic way.

Lastly, we briefly discuss the experimental feasibility.
Due to rich manipulation techniques, such as individual site control and large range of tunable parameters, our protocols can be realized via superconducting quantum circuits~\cite{AAHouckNP2012, MHDevoretScience2013, YChenPRL2014, ABaustPRB2015, XHDengPRB2015}.
It has been demonstrated that by controlling the flux threading the non-hysteretic rf SQUID loop between the two resonators, the coupling strength between two superconducting transmission line resonators can be tuned from negative to positive~\cite{FWulschnerEPJ2016}.
This experimental technique enables the desired tunneling modulation in our system.
Moreover, one may use the photonic lattices with complex couplings as a potential platform for testing our protocols~\cite{ABellOptica2017}.

\acknowledgements{This work is supported by the National Natural Science Foundation of China (NNSFC) under Grants Grants No. 11874434 and 11574405. Y. K. was partially supported by International Postdoctoral Exchange Fellowship Program (No. 20180052).}

\setcounter{equation}{0}
\renewcommand{\theequation}{A\arabic{equation}}

\section*{APPENDIX A: Derivation of the Eq.~(4)}
Applying the position operator $\hat{X}$ on the Wannier state, we have
\begin{widetext}
\begin{eqnarray}\label{Eq.XWannier}
\hat{X}|W_{m}(R,0)\rangle
&=&\frac{1}{L}\sum_{k,j}e^{-ikqR}e^{ikj}u_{m,j}(k,0)jc_{j}^{\dagger}|0\rangle\cr\cr
&=&qR|W_{m}(R,0)\rangle+\frac{1}{L}\sum_{k,j}
e^{-ikqR}e^{ikj}i\frac{\partial}{\partial k}u_{m,j}(k,0)
c_{j}^{\dagger}|0\rangle.
\end{eqnarray}
Here we use the relation
$\frac{\partial}{\partial k}|W_{m}(R,t)\rangle=0$.
Thus the mean position at $t=0$ is given as
\begin{eqnarray}\label{Eq.Xmean0A}
\langle\hat{X}_{m}(0)\rangle
&=&qR+\frac{1}{L^{2}}\sum\limits_{k,k',j}e^{-i(k-k')qR}e^{i(k-k')j}
u^{\ast}_{m,j}(k',0)
i\frac{\partial}{\partial k}u_{m,j}(k,0)\cr\cr
&=& qR+\frac{1}{L^{2}}\sum\limits_{k,k'}e^{-i(k-k')qR}
\sum\limits_{R'=0}^{L-1}\sum\limits_{j'=1}^q e^{i(k-k')(qR'+j')}u^{\ast}_{m,j'}(k',0)
i\frac{\partial}{\partial k}u_{m,j'}(k,0)\cr\cr
&=& qR+\frac{1}{L}\sum\limits_{k}\sum\limits_{j'=1}^q
u^{\ast}_{m,j'}(k,0)
i\frac{\partial}{\partial k}u_{m,j'}(k,0)\cr\cr
&=&qR+\frac{1}{L}\sum_{k}
\langle u_{m}(k,0)|
i\frac{\partial}{\partial k}|u_{m}(k,0)\rangle,
\end{eqnarray}
%
\section*{APPENDIX B: Derivation of the Eq.~(16)}
We show how to derive the Eq.~(16).
To do this, we calculate the element
$\langle W_{m}(R,T)|\hat{X}|W_{m}(0,T)\rangle$ and obtain
\begin{eqnarray}\label{Eq.WmRTXWm0T}
&&\langle W_{m}(R,T)|\hat{X}|W_{m}(0,T)\rangle\cr\cr
&&=\frac{1}{L}\sum_{k,k',j}e^{ik'qR}e^{i(\gamma(k)-\gamma(k'))} e^{i(k-k')j} u_{m,j}^{\ast}(k',0)
\Big[\Big(-\frac{\partial}{\partial k}\gamma(k)\Big)u_{m,j}(k,0)
+i\frac{\partial}{\partial k}u_{m,j}(k,0)\Big]\cr\cr
&&=\frac{1}{L}\sum_{k}e^{ikqR}
\big(\langle u_{m}(k,0)|i\frac{\partial}{\partial k}|u_{m}(k,0)\rangle
-\frac{\partial}{\partial k}\gamma(k)\big),
\end{eqnarray}
The derivation of Eq.~\eqref{Eq.WmRTXWm0T} is similar to Eq.~\eqref{Eq.Xmean0}.
Thus we get
\begin{eqnarray}\label{Eq.OmegaDTA}
\Omega_{D}(T)
&=&\sum_{R}|\langle W_{m}(R,T)|\hat{X}|W_{m}(0,T)\rangle|^{2}
-|\langle W_{m}(0,T)|\hat{X}|W_{m}(0,T)\rangle|^{2}\cr\cr
&=&\frac{1}{L^{2}}\sum_{R,k,k'}e^{i(k-k')qR}
\Big(\langle u_{m}(k,0)|i\frac{\partial}{\partial k}|u_{m}(k,0)\rangle
-\frac{\partial}{\partial k}\gamma(k)\Big)\times\cr\cr
&&\Big(\langle u_{m}(k',0)|i\frac{\partial}
{\partial k'}|u_{m}(k',0)\rangle
-\frac{\partial}{\partial k'}\gamma(k')\Big)
-\Big(\frac{1}{L}\sum_{k}\langle u_{m}(k,0)|i\frac{\partial}
{\partial k}|u_{m}(k,0)\rangle-\frac{\partial}{\partial k}\gamma(k)
\Big)^{2}\cr\cr
&=&\frac{1}{L}\sum_{k}\Big(\langle u_{m}(k,0)|i\frac{\partial}
{\partial k}|u_{m}(k,0)\rangle-\frac{\partial}{\partial k}\gamma(k)
\Big)^{2}
-\Big(\frac{1}{L}\sum_{k}\langle u_{m}(k,0)|i\frac{\partial}
{\partial k}|u_{m}(k,0)\rangle-\frac{\partial}{\partial k}\gamma(k)
\Big)^{2}\cr\cr
&=&\frac{1}{L}\sum_{k}\Big[\Big(\langle u_{m}(k,0)|i\frac{\partial}
{\partial k}|u_{m}(k,0)\rangle-\frac{\partial}{\partial k}\gamma(k)\Big)
-\Big(\frac{1}{L}\sum_{k}\langle u_{m}(k,0)|i\frac{\partial}
{\partial k}|u_{m}(k,0)\rangle-\frac{\partial}{\partial k}\gamma(k)\Big)\Big]^{2}\cr\cr
&=&\frac{1}{L}\sum_{k}\Big(-\frac{\partial}{\partial k}\gamma(k)
+\frac{1}{L}\sum_{k}\frac{\partial}{\partial k}\gamma(k)\Big)^{2}.
\end{eqnarray}
Here, we use the relation
\begin{equation}
\langle u_{m}(k,0)|i\frac{\partial}{\partial k}|u_{m}(k,0)\rangle=
\frac{1}{L}\sum\limits_k \langle u_{m}(k,0)|
i\frac{\partial}{\partial k}|u_{m}(k,0)\rangle,
\end{equation}
for the MLWSs~\cite{NMarzariPRB1997,NMarzariRMP2012,YKePRA2017}.
\end{widetext}



\begin{thebibliography}{99}

\bibitem{DJThoulessPRB1983}
D. J. Thouless, Quantization of particle transport, Phys. Rev.
B \textbf{27}, 6083 (1983).
%

\bibitem{LWangPRL2013}
L. Wang, M. Troyer, and X. Dai, Topological Charge Pumping
in a One-Dimensional Optical Lattice, Phys. Rev. Lett. \textbf{111}, 026802 (2013).
%

\bibitem{YKeLPR2016}
Y. Ke, X. Qin, F. Mei, H. Zhong, Y. S. Kivshar, and C. Lee,
Topological phase transitions and Thouless pumping of light
in photonic waveguide arrays, Laser Photon. Rev. \textbf{10}, 995 (2016).
%

\bibitem{MLohseNP2016}
M. Lohse, C. Schweizer, O. Zilberberg, M. Aidelsburger, and I.
Bloch, A Thouless quantum pump with ultracold bosonic atoms
in an optical superlattice, Nat. Phys. \textbf{12}, 350 (2016).
%

\bibitem{SNakajimaNP2016}
S. Nakajima, T. Tomita, S. Taie, T. Ichinose, H. Ozawa, L. Wang,
M. Troyer, and Y. Takahashi, Topological Thouless pumping of
ultracold fermions, Nat. Phys. \textbf{12}, 296 (2016).
%

\bibitem{JTangpanitanonPRL2016}
J. Tangpanitanon, V. M. Bastidas, S. Al-Assam, P. Roushan,
D. Jaksch, and D. G. Angelakis, Topological Pumping of Photons in Nonlinear Resonator Arrays, Phys. Rev. Lett. \textbf{117}, 213603 (2016)
%

\bibitem{YKePRA2017}
Y. Ke, X. Qin, Y. S. Kivshar, and C. Lee, Multiparticle Wannier states and Thouless pumping of interacting bosons, Phys. Rev. A \textbf{95}, 063630 (2017).
%

\bibitem{AHaywardPRB2018}
A. Hayward, C. Schweizer, M. Lohse, M. Aidelsburger, and F. Heidrich-Meisner, Topological charge pumping in the interacting bosonic Rice-Mele model, Phys. Rev. B \textbf{98}, 245148 (2018).
%

\bibitem{MLohseNature2018}
M. Lohse, C. Schweizer, H. M. Price, O. Zilberberg, and I. Bloch, Exploring 4D quantum Hall physics with a 2D topological charge pump, Nature (London) \textbf{553}, 55 (2018).
%

\bibitem{IMartinPRX2017}
I. Martin, G. Refael, and B. Halperin, Topological Frequency Conversion in Strongly Driven Quantum Systems, Phys. Rev. X \textbf{7}, 041008 (2017).
%

\bibitem{PWeinbergPR2017}
P. Weinberg, M. Bukov, L. D'Alessio, A. Polkovnikov,
S. Vajna, and M. Kolodrubetz, Adiabatic perturbation theory and geometry of periodically-driven systems, Phys. Rep. \textbf{688}, 1 (2017).
%

\bibitem{MHKolodrubetzPRL2018}
M. H. Kolodrubetz, F. Nathan, S. Gazit, T. Morimoto, and J. E. Moore, Topological Floquet-Thouless Energy Pump, Phys. Rev. Lett. \textbf{120}, 150601 (2018).
%

\bibitem{WMaPRL2018}
W. Ma, L. Zhou, Q. Zhang, M. Li, C. Cheng, J. Geng, X. Rong,
F. Shi, J. Gong, and J. Du, Experimental Observation of a Generalized Thouless Pump with a Single Spin, Phys. Rev. Lett. \textbf{120}, 120501 (2018).
%

\bibitem{LPriviteraPRL2018}
L. Privitera, A. Russomanno, R. Citro, and G. E. Santoro, Nonadiabatic Breaking of Topological Pumping, Phys. Rev. Lett. \textbf{120}, 106601 (2018).
%

\bibitem{YKunoarXiv2018}
Yoshihito Kuno, Non-adiabatic extension of the Zak phase and charge pumping in the Rice-Mele model, arXiv 1809.05702 (2018).
%

\bibitem{NLang2017}
N. Lang and H. P. Buchle, Topological networks for quantum communication between distant qubits, npj Quantum Information \textbf{3}, 47 (2017).
%

\bibitem{CDlaskaQSC2017}
C. Dlaska, B. Vermersch, and P. Zoller, Robust quantum state transfer via topologically protected edge channels in dipolar arrays, Quantum Sci. Technol. \textbf{2}, 015001 (2017).
%

\bibitem{ABRedondoScience2018}
A. B. Redondo, B. Bell, D. Oren, B. J. Eggleton, and M. Segev, Topological protection of biphoton states, Science \textbf{362}, 568 (2018).
%

\bibitem{ELHahnPR1950}
E. L. Hahn, Spin Echoes, Phys. Rev. \textbf{80}, 580 (1950).
%

\bibitem{LMKVandersypenRMP2004}
L. M. K. Vandersypen and I. L. Chuang, NMR techniques for quantum control and computation, Rev. Mod. Phys. \textbf{76}, 1037 (2004).
%

\bibitem{MAtalaNP2013}
M. Atala, M. Aidelsburger, J. T. Barreiro, D. Abanin, T. Kitagawa,
E. Demler, and I. Bloch, Direct measurement of the Zak phase in
topological Bloch bands, Nat. Phys. \textbf{9}, 795 (2013).
%

\bibitem{DSuterRMP2016}
D. Suter and G. A. ¨¢lvarez, Colloquium: Protecting quantum information against environmental noise, Rev. Mod. Phys. \textbf{88}, 041001 (2016).
%

\bibitem{MGarttnerNP2017}
M. G\"{a}rttner, J. G. Bohnet, A. Safavi-Naini, M. L. Wall, J. J. Bollinger, and A. M. Rey, Measuring out-of-time-order correlations and
multiple quantum spectra in a trapped-ion quantum magnet, Nat. Phys. \textbf{13}, 781 (2017).
%

\bibitem{PGHarperPPSA1955}
P. G. Harper, Single Band Motion of Conduction Electrons in a
Uniform Magnetic Field, Proc. Phys. Soc. A \textbf{68}, 874 (1955).
%

\bibitem{SAubryAIPS1980}
S. Aubry and G. Andr\'{e}, Ann. Isr. Phys. Soc. \textbf{3}, 133 (1980).
%

\bibitem{LLangPRL2012}
L. Lang, X. Cai, and S. Chen, Edge States and Topological Phases in One-Dimensional Optical Superlattices, Phys. Rev. Lett. \textbf{108}, 220401 (2012).
%

\bibitem{GHWannierPR1937}
G. H. Wannier, The Structure of Electronic Excitation Levels in
Insulating Crystals, Phys. Rev. \textbf{52}, 191 (1937).
%

\bibitem{NMarzariPRB1997}
N. Marzari and D. Vanderbilt, Maximally localized generalized
Wannier functions for composite energy bands, Phys. Rev. B \textbf{56}, 12847 (1997).
%

\bibitem{NMarzariRMP2012}
N. Marzari, A. A. Mostofi, J. R. Yates, I. Souza, and D.
Vanderbilt, Maximally localized Wannier functions: Theory and
applications, Rev. Mod. Phys. \textbf{84}, 1419 (2012).
%

\bibitem{RDKingPRB1993}
R. D. King-Smith and D. Vanderbilt, Theory of polarization of
crystalline solids, Phys. Rev. B \textbf{47}, 1651 (1993).
%

\bibitem{DXiaoRMP2010}
D. Xiao, M.-C. Chang, and Q. Niu, Berry phase effects on electronic properties, Rev. Mod. Phys. \textbf{82}, 1959 (2010).
%

\bibitem{MTakahashiJPC1977}
M. Takahashi, Half-filled Hubbard model at low temperature, J. Phys. C: Solid State Phys. \textbf{10}, 1289 (1977).
%

\bibitem{SBravyiAP2011}
S. Bravyi, D. P. DiVincenzo, and D. Loss, Schrieffer¨CWolff transformation for quantum many-body systems, Annals of Physics \textbf{326}, 2793 (2011).
%

\bibitem{KMizutaPRL2018}
K. Mizuta, K. Takasan, M. Nakagawa, and N. Kawakami, Spatial-Translation-Induced Discrete Time Crystals, Phys. Rev. Lett. \textbf{121}, 093001 (2018).
%

\bibitem{PTitumPRX2016}
P. Titum, E. Berg, M. S. Rudner, G. Refael, and N. H. Lindner, Anomalous Floquet-Anderson Insulator as a Nonadiabatic Quantized Charge Pump, Phys. Rev. X \textbf{6}, 021063 (2016).
%

\bibitem{AAHouckNP2012}
A. A. Houck, H. E. Tureci, and J. Koch, On-chip quantum simulation with superconducting circuits, Nat. Phys. \textbf{8}, 292 (2012).
%

\bibitem{MHDevoretScience2013}
M. H. Devoret and R. J. Schoelkopf, Superconducting circuits for quantum information: an outlook, Science \textbf{339}, 1169 (2013).
%

\bibitem{YChenPRL2014}
Y. Chen, C. Neill, P. Roushan, N. Leung, M. Fang, R. Barends, J. Kelly, B. Campbell, Z. Chen, B. Chiaro, A. Dunsworth, E. Jeffrey, A. Megrant, J. Y. Mutus, P. J. J. O¡¯Malley, C. M. Quintana, D. Sank, A. Vainsencher, J. Wenner, T. C. White, M. R. Geller, A. N. Cleland, and J. M. Martinis, Qubit architecture with high coherence and fast tunable coupling, Phys. Rev. Let. \textbf{113}, 220502 (2014).
%

\bibitem{ABaustPRB2015}
A. Baust, E. Hoffmann, M. Haeberlein, M. J. Schwarz, P. Eder, J. Goetz, F. Wulschner, E. Xie, L. Zhong, F. Quijandr\'{i}a, B. Peropadre, D. Zueco, J.-J. G. Ripoll, E. Solano, K. Fedorov, E. P. Menzel, F. Deppe, A. Marx, and R. Gross, Tunable and switchable coupling between two superconducting resonators, Phys. Rev. B \textbf{91}, 014515 (2015).
%

\bibitem{XHDengPRB2015}
X.-H. Deng, C.-J. Jia, and C.-C. Chien, Sitewise manipulations and Mott insulator-superfluid transition of interacting photons using superconducting circuit simulators, Phys. Rev. B \textbf{91}, 054515 (2015).
%

\bibitem{FWulschnerEPJ2016}
F. Wulschner, J. Goetz, F. R Koessel, E. Hoffmann, A. Baust, P. Eder, M. Fischer, M. Haeberlein, M. J. Schwarz, M. Pernpeintner, E. Xie, L. Zhong, C. W. Zollitsch, B. Peropadre, J.-J. G. Ripoll, E. Solano, K. G. Fedorov, E. P. Menzel, F. Deppe, A. Marx, and R. Gross, Tunable coupling of transmission-line microwave resonators mediated by an rf SQUID, EPJ Quant. Technol. \textbf{3}, 10 (2016).
%

\bibitem{ABellOptica2017}
A. Bell, K. Wang, S. Solntsev, N. Neshev, A. Sukhorukov, and J. Eggleton, Spectral photonic lattices with complex long-range coupling, Optica. \textbf{4}, 1433 (2017).

\end{thebibliography}
\end{document}